\documentclass[modern]{aastex63}

\usepackage{amsmath,mathtools,mathrsfs,amssymb}

\shorttitle{Resolving supermassive black hole binaries}
\graphicspath{{./}{figures/}}

\begin{document}

\title{Determining sub-parsec supermassive black hole binary orbits with infrared interferometry}

\email{jason.dexter@colorado.edu}

\author[0000-0003-3903-0373]{Jason Dexter}
\affiliation{JILA and Department of Astrophysical and Planetary Sciences, University of Colorado, Boulder, CO 80309, USA}

\author[0000-0003-0291-9582]{Dieter Lutz}
\affiliation{Max-Planck-Institut f\"ur Extraterrestrische Physik, Giessenbachstr. 1, D-85748 Garching, Germany}

\author[0000-0002-2125-4670]{T. Taro Shimizu}
\affiliation{Max-Planck-Institut f\"ur Extraterrestrische Physik, Giessenbachstr. 1, D-85748 Garching, Germany}

\author[0000-0002-4569-9009]{Jinyi Shangguan}
\affiliation{Max-Planck-Institut f\"ur Extraterrestrische Physik, Giessenbachstr. 1, D-85748 Garching, Germany}

\author[0000-0003-4949-7217]{Richard I. Davies}
\affiliation{Max-Planck-Institut f\"ur Extraterrestrische Physik, Giessenbachstr. 1, D-85748 Garching, Germany}

\author[0000-0003-4175-3474]{P. Tim de Zeeuw}
\affiliation{Sterrewacht Leiden, Leiden University, Postbus 9513, 2300 RA Leiden, The Netherlands}
\affiliation{Max-Planck-Institut f\"ur Extraterrestrische Physik, Giessenbachstr. 1, D-85748 Garching, Germany}

\author[0000-0002-0018-3666]{Eckhard Sturm}
\affiliation{Max-Planck-Institut f\"ur Extraterrestrische Physik, Giessenbachstr. 1, D-85748 Garching, Germany}

\author{Frank Eisenhauer}
\affiliation{Max-Planck-Institut f\"ur Extraterrestrische Physik, Giessenbachstr. 1, D-85748 Garching, Germany}

\author{Natascha M. F\"{o}rster-Schreiber}
\affiliation{Max-Planck-Institut f\"ur Extraterrestrische Physik, Giessenbachstr. 1, D-85748 Garching, Germany}

\author[0000-0002-2581-9114]{Feng Gao}
\affiliation{Max-Planck-Institut f\"ur Extraterrestrische Physik, Giessenbachstr. 1, D-85748 Garching, Germany}

\author[0000-0002-2767-9653]{Reinhard Genzel}
\affiliation{Max-Planck-Institut f\"ur Extraterrestrische Physik, Giessenbachstr. 1, D-85748 Garching, Germany}
\affiliation{Departments of Physics and Astronomy, Le Conte Hall, University of California, Berkeley, CA 94720, USA}

\author[0000-0002-5708-0481]{Stefan Gillessen}
\affiliation{Max-Planck-Institut f\"ur Extraterrestrische Physik, Giessenbachstr. 1, D-85748 Garching, Germany}

\author{Oliver Pfuhl}
\affiliation{European Southern Observatory, Karl-Schwarzschild-Str. 2, 85748 Garching, Germany}

\author{Linda J. Tacconi}
\affiliation{Max-Planck-Institut f\"ur Extraterrestrische Physik, Giessenbachstr. 1, D-85748 Garching, Germany}

\author[0000-0002-0327-6585]{Felix Widmann}
\affiliation{Max-Planck-Institut f\"ur Extraterrestrische Physik, Giessenbachstr. 1, D-85748 Garching, Germany}

\begin{abstract}
  Radial velocity monitoring has revealed the presence of moving broad emission lines in some quasars, potentially indicating the presence of a sub-parsec binary system. Phase-referenced, near-infrared interferometric observations could map out the binary orbit by measuring the photocenter difference between a broad emission line and the hot dust continuum. We show that astrometric data over several years may be able to detect proper motions and accelerations, confirming the presence of a binary and constraining system parameters. The brightness, redshifts, and astrometric sizes of current candidates are well matched to the capabilities of the upgraded VLTI/GRAVITY+ instrument, and we identify a first sample of $10$ possible candidates. The astrometric signature depends on the morphology and evolution of hot dust emission in supermassive black hole binary systems. Measurements of the photocenter offset may reveal binary motion whether the hot dust emission region is fixed to the inner edge of the circumbinary disk, or moves in response to the changing irradiation pattern from an accreting secondary black hole.\\
\\
\end{abstract}


\section{Introduction} \label{sec:intro}

Central supermassive black holes in merging galaxies are thought to be efficiently driven to $\lesssim 10$ pc separations by dynamical friction \citep{begelman1980}. Their further evolution remains uncertain. Interactions with gas in a circumbinary accretion disk could either drive the binary closer together \citep[][]{armitage2002} or further apart \citep[e.g.,][]{munoz2019}. Detections of sub-pc supermassive black hole binaries (SMBHBs) would provide important input to galaxy formation models \citep[][]{volonteri2003}, estimates of the stochastic gravitational wave background \citep[e.g.,][]{siemens2013}, and the rate of individual merger events seen by LISA \citep[e.g.,][]{amaro2012}. 

Growing numbers of dual active galactic nuclei (AGN) are seen on kpc scales in interacting or post-merger galaxies \citep[][]{comerford2009}. The closest known supermassive black hole pair has a projected separation of $\simeq 7$ pc  \citep{rodriguez2006}, detected with radio very long baseline interferometry. Suggested evidence of sub-pc binaries comes from AGN with double-peaked broad emission lines \citep[][]{gaskell1983}, offset and moving broad emission lines \citep[][]{eracleous2012}, and periodically varying optical light curves \citep[][]{graham2015}.

Infrared interferometry with the VLT Interferometer instrument GRAVITY \citep{gravity2017firstlight} can now spatially resolve the broad emission line region (BLR) in the brightest AGN on sky by measuring its velocity-dependent photocenter offset from the hot dust continuum \citep{gravity2018}. For a system with double-peaked broad lines, an extension of this method could reveal the presence of an SMBHB \citep{songsheng2019}. Several candidate double-peaked systems have been ruled out as binaries  \citep{eracleous1997,decarli2013}, and both black holes are only expected to be actively accreting and retain their individual BLRs over a narrow region of parameter space  \citep[][]{bogdanovic2008,shen2010}.

Monitoring campaigns have identified a number of candidates with single-peaked, offset, and moving emission lines  \citep{runnoe2017,guo2019}. Here we consider the requirements for astrometrically confirming the presence of a binary in these systems. Over a relevant range of parameter space, relative astrometry between the BLR of an accreting secondary black hole and hot dust in the surrounding circumbinary disk could map out the binary orbit (\autoref{sec:param_space}). The observational requirements, given the current candidate systems, are well matched to the sensitivity of the planned upgrade of the GRAVITY instrument, GRAVITY+ (\autoref{sec:gplus}). A monitoring campaign over $\simeq 5-10$ years could be sufficient to detect both proper motion and acceleration in these systems, constraining the system parameters and potentially providing robust detections of sub-pc SMBHBs. Possible  extensions of this study including the prospects of additional measurements and targets are discussed in \autoref{sec:discussion}.\\

\section{Astrometric mapping of supermassive black hole binaries}
\label{sec:param_space}

We assume a binary system of total mass $M = M_1 + M_2$ and mass ratio of $q = M_2/M_1 \le 1$ in a circular orbit. The orbital period and semi-major axis on sky are then,

\begin{align}
    P  \simeq 95 \left(\frac{a}{0.1 \, \rm pc}\right)^{3/2} \left(\frac{M}{10^9 M_\sun}\right)^{-1/2} \rm yr,\\
    \theta_a \simeq 34 \left(\frac{a}{0.1 \, \rm pc}\right) \left(\frac{D_A}{600 \, \rm Mpc}\right)^{-1} \mu \rm as,
\end{align}

\noindent where $D_A$ is the source angular diameter distance. We further assume that the SMBHB is surrounded by a circumbinary gas disk, which is centered on the system center of mass and truncated at a radius $\simeq 2 a$ \citep{artymowicz1994}. Accretion proceeds through a central, low density cavity via thin streams, forming ``mini-disks" around the two black holes \citep[e.g.,][]{cuadra2009,noble2012,dorazio2013,bowen2018}.

\begin{figure}[t]
\begin{tabular}{ll}
\includegraphics[width=0.46\textwidth]{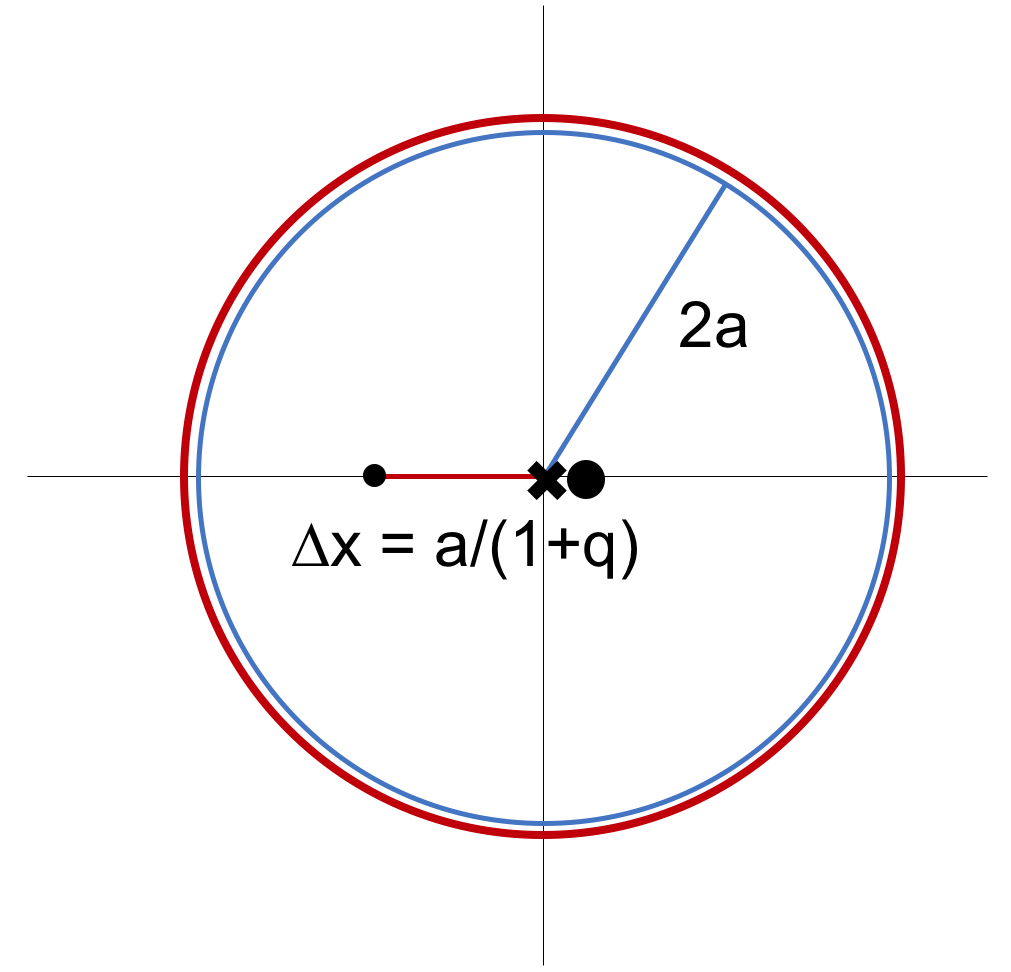} &
\includegraphics[width=0.46\textwidth]{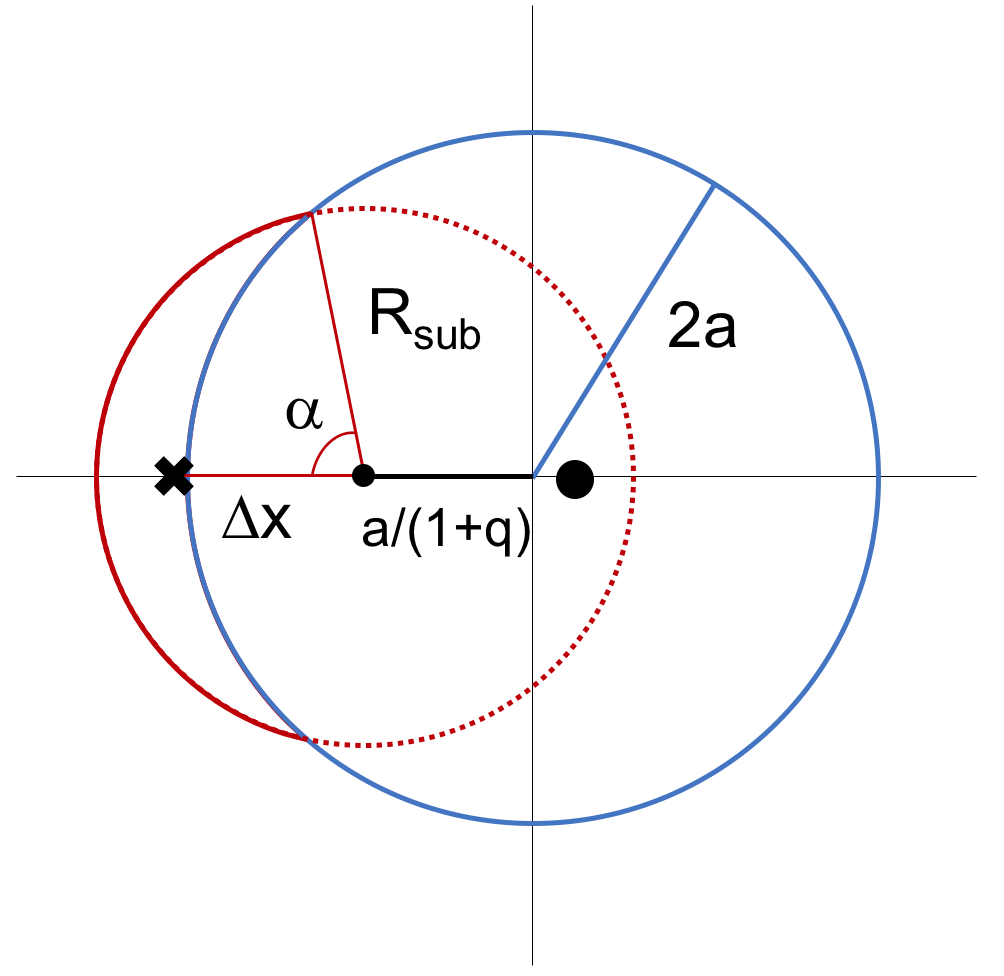}
\end{tabular}
\caption{Geometry of the static (left) and evolving (right) models for the continuum hot dust emission. In both panels, the primary and secondary black holes are shown as filled black points. In the static scenario, the hot dust photocenter (black cross) is assumed to be fixed at the binary center of mass, e.g. as the result of an emission region (thick, red circle) concentrated near the inner edge of the circumbinary disk (thin blue circle), whose center coincides with the center of mass. In the evolving dust scenario, we calculate the continuum photocenter as the centroid (black cross) of the shaded red arc of half-angle $\alpha$ where the sublimation radius $R_{\rm sub}$ lies inside the circumbinary disk. The offset $\Delta x$ is the line segment between the center of the red circle and the cross.}
\label{fig:circle_diagram}
\end{figure}

\begin{figure}[t]
\centering
\includegraphics[width=0.5\textwidth]{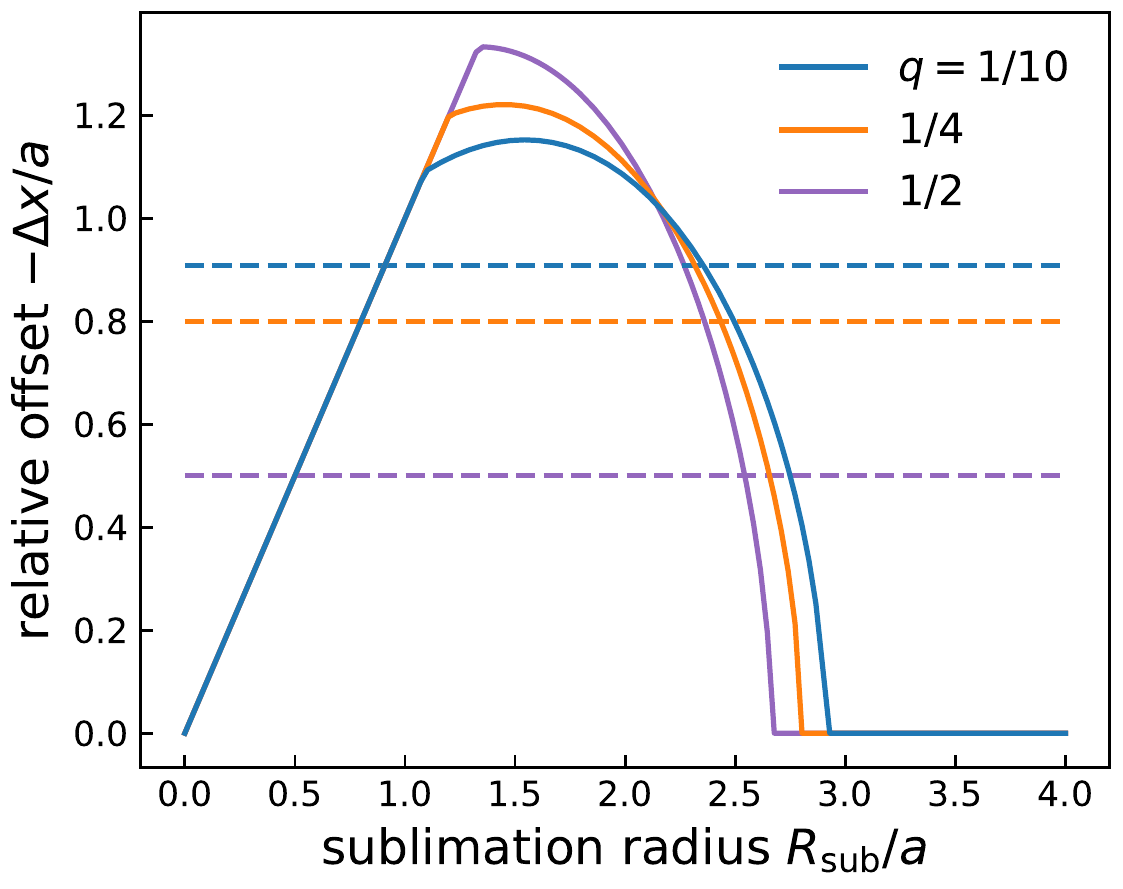}
\caption{Astrometric offset for the geometric evolving continuum model as a function of sublimation radius $R_{\rm sub}$ for three values of $q$, both measured in units of the orbital semi-major axis $a$. When the circle of radius $R_{\rm sub}$ centered on the secondary is fully inside the cavity ($R_{\rm sub}/a < (1+2q)/(1+q)$), the offset $-\Delta x = R_{\rm sub}$ (linear rise). When the circle partially intersects the circumbinary disk, the offset is calculated as the centroid of the arc lying inside the circumbinary disk. The offset vanishes once the circle lies entirely in the circumbinary disk and the dust emission is assumed to be centered on the secondary. The $R_{\rm sub}/a > 2.5$ limit is not encountered in practice, since the condition that the BLR is bound to the secondary is more constraining. The dashed lines show the magnitude (with opposite sign) of the astrometric offset of the secondary from the center of mass in each case, $\Delta x = a/(1+q)$. We find similar astrometric amplitudes and evolution in both scenarios for $0.5 \lesssim R_{\rm sub}/a \lesssim 2.5$.}
\label{fig:offset_circle}
\end{figure}

\begin{figure*}[t]
\begin{tabular}{cc}
\includegraphics[width=0.47\textwidth]{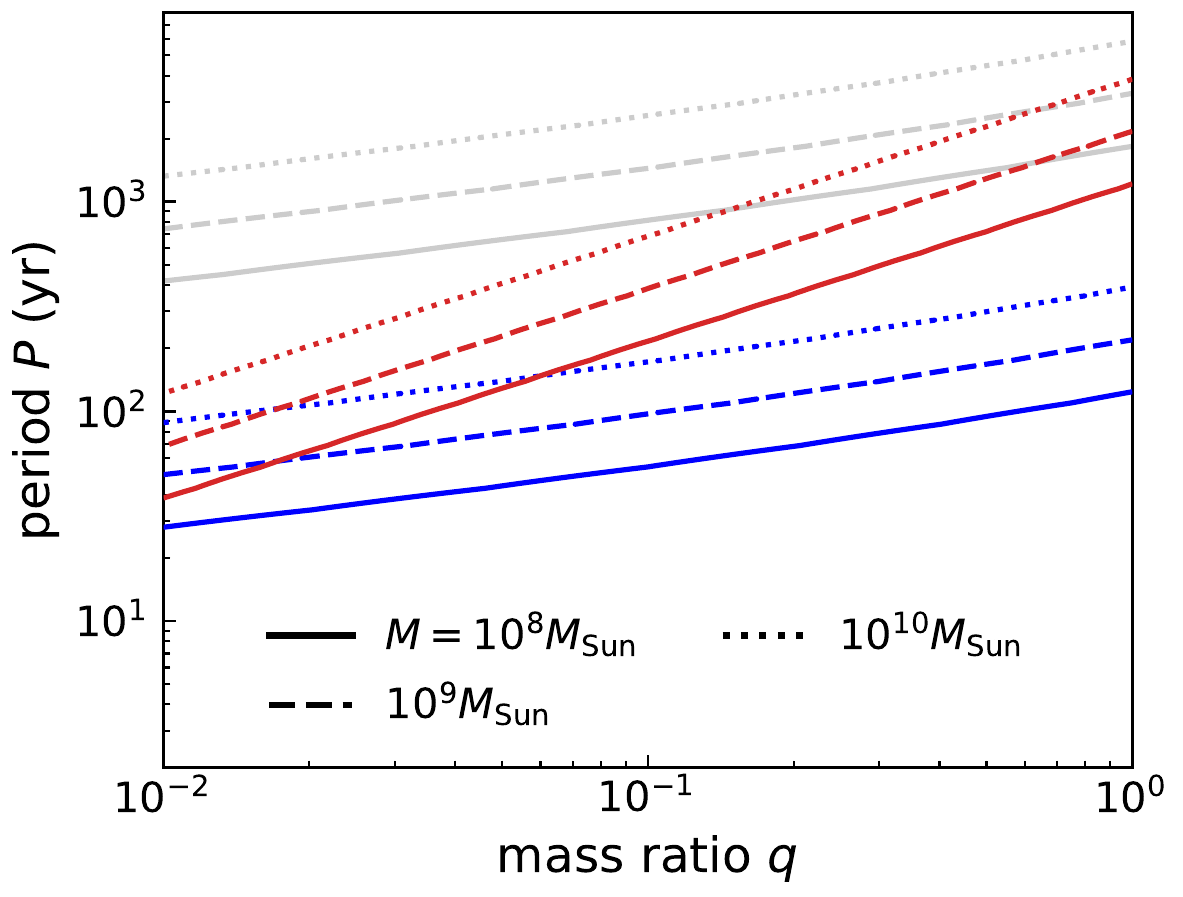} &
\includegraphics[width=0.47\textwidth]{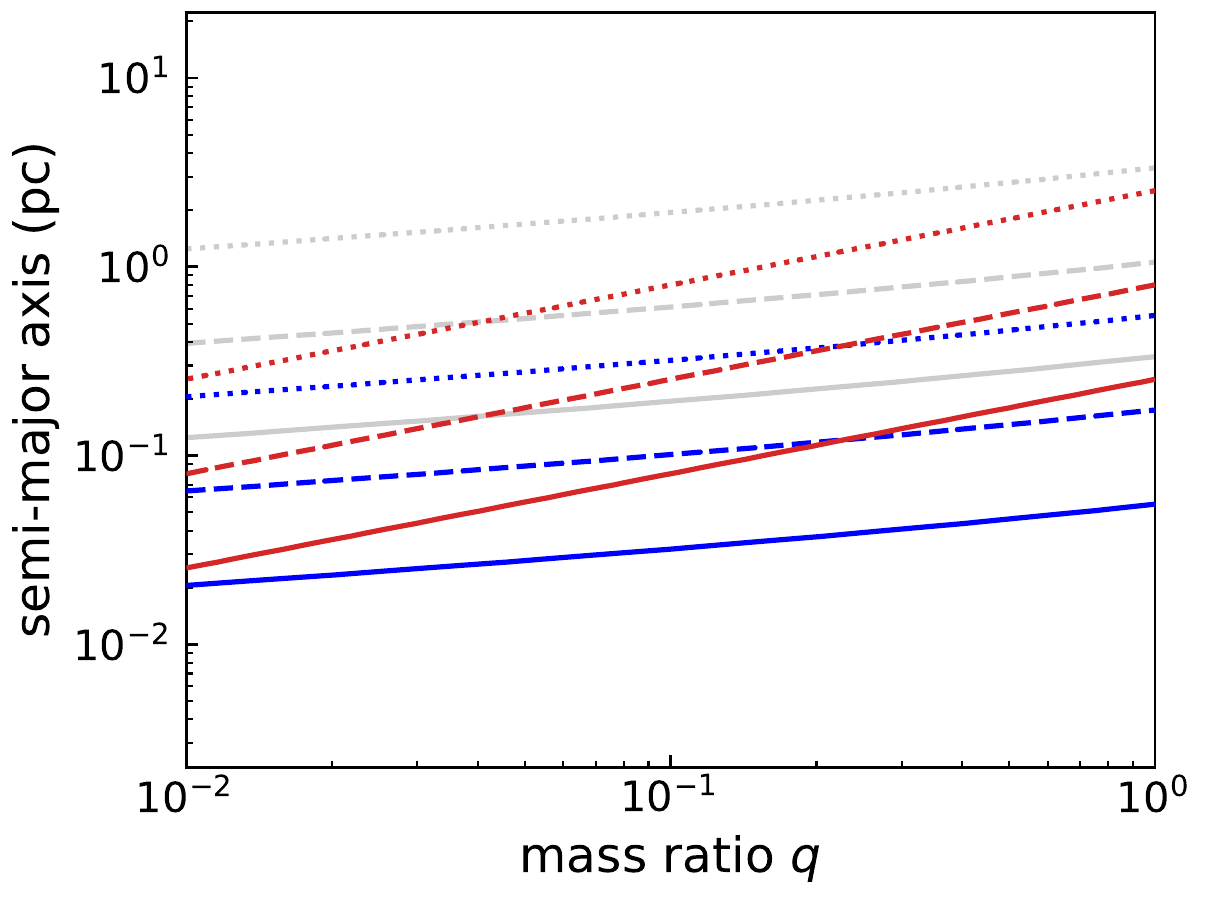}
\end{tabular}
\caption{Contours of $R_{\rm BLR} = R_L$ (blue), $R_{\rm sub} = R_L$ (light gray), and $0.5 \le R_{\rm sub}/a \le 2.5$ (dark red) as a function of period $P$ (left) or semi-major axis $a$ (right) and mass ratio $q$ for total masses of $M = 10^8$, $10^9$, and $10^{10} M_\sun$. The BLR and sublimation radii depend on luminosity, $L \propto q M$. In all cases, the region of interest for spectro-astrometry would be above the blue curves and below the light gray curves for each bin in $M$, where the BLR would remain bound to the accreting secondary black hole while the hot dust would not. The more restrictive parameter space below the dark red lines shows where the astrometric offset of the two components could be used to trace the SMBHB orbit even if the hot dust continuum emission follows the secondary's orbit at a distance of $R_{\rm sub}$.}
\label{fig:paramspace}
\end{figure*}

\subsection{Relevant parameter regime}

Mapping out the binary orbit requires an astrometric measurement of a light source centered on one of the black holes. With near-infrared observations, the most promising candidate is a broad emission line from ionized gas bound to one of the black holes. For concreteness, we assume that this is the secondary black hole $M_2$. Many simulations find a much higher accretion rate onto the secondary \citep[e.g.,][]{cuadra2009,dorazio2013,munoz2019,duffell2020}. This is also the assumption made by recent radial velocity studies \citep[e.g.,][]{runnoe2017}, allowing for a direct comparison. With infrared interferometry, we also need a reference source. Here we consider the method in current use, where the broad emission line is phase-referenced to the continuum emission radiated by the surrounding hot dust. 

Two requirements to make this measurement are that 1) the BLR is bound to the secondary black hole ($R_{\rm BLR} < R_L$, where $R_L$ is the Roche-Lobe radius as approximated by \citealt{eggleton1983}), while 2) hot dust is not ($R_{\rm sub} > R_L$, where $R_{\rm sub}$ is the sublimation radius). We estimate $R_{\rm BLR} \simeq 0.07 L_{2, 46}$ pc and $R_{\rm sub} \simeq 0.4 L_{2, 46}$ pc using scaling relations with luminosity as measured separately for the BLR \citep[e.g.,][]{bentz2013} and near-infrared continuum \citep{suganuma2006,kishimoto2011,gravity2020dust}. The luminosity of the secondary black hole is $L_2 = \epsilon L_{\rm Edd} (q M)$, where $L_{\rm Edd}$ is the Eddington luminosity and $\epsilon = 0.1$ is the assumed Eddington ratio of the secondary. Any viable candidates identified by the radial velocity method would by definition have a BLR bound to the secondary. Even large graphite grains, often assumed responsible for the NIR continuum \citep[e.g.,][]{kishimoto2007}, should be sublimated within the Roche-Lobe of the secondary for binary orbital periods of $\lesssim 10^3$ yr {(below the gray lines in \autoref{fig:paramspace}). 

The major uncertainty in this scenario is where the near-infrared continuum emission originates, and how it evolves over the course of a binary orbit. We consider two scenarios (\autoref{fig:circle_diagram}). (i) If the continuum emission is stationary, e.g., tracing the inner edge of the circumbinary disk, then relative astrometry of the BLR measures the secondary's orbit. (ii) Empirically, the near-infrared emission size scales with that expected for the sublimation radius. It seems possible that the continuum emission could instead preferentially originate in the regions of the circumbinary disk closest to the secondary, where the irradiating flux is strongest and dust temperatures highest. In that case, both the line and continuum emission could track the binary orbit, although we have not tested this using radiative transfer calculations including dust heating, anisotropic emission, or obscuration along the line of sight.

We have developed a simple geometric model for the second ``evolving continuum" scenario. Hot dust is assumed to form outside the binary and at the sublimation radius of the secondary. The possible emission locations are then along a circle of radius $R_{\rm sub}$ centered on the position of the secondary. When the sublimation radius intersects the circumbinary disk, we assume that hot dust emission is produced with equal intensity everywhere along the circle where it intersects the circumbinary disk. When the sublimation radius is smaller than the distance from the secondary to the edge of the circumbinary disk, we assume that some small region (e.g., in an accretion stream) at a distance of $\simeq R_{\rm sub}$ will form and radiate hot dust instead. The astrometric shift is then the offset between the secondary black hole and the continuum photocenter.

The expression is derived in \autoref{app:circles} and the result is shown in \autoref{fig:offset_circle}. At very small $R_{\rm sub}/a$ the offset is small because hot dust forms close to the secondary. Once $R_{\rm sub}$ becomes large enough to heat dust all along the circumbinary disk, the continuum photocenter is at the position of the secondary black hole, and the astrometric shift vanishes. For a range of $0.5 \lesssim R_{\rm sub}/a \lesssim 2.5$, the relative offset is similar in magnitude to the true orbital offset. We plot this parameter space constraint as the dark red lines in \autoref{fig:paramspace}. It is more restrictive than simply requiring that hot dust cannot be bound to the secondary. In particular, for $q \ll 1$ the available parameter space shrinks until a minimum $q_{\rm min} \simeq 6 \times 10^{-3}$ where no solutions are possible. Still, the geometric model suggests that relative astrometry might trace the binary orbit over much of the relevant parameter space, even if the near-infrared continuum is tracking the motion of the secondary.

\subsection{Supermassive black hole binary astrometry}

We next consider the radial velocity and astrometric position of the secondary black hole on sky. Following \citet{eracleous2012}, we write the radial velocity as,

\begin{equation}
\label{eq:rv}
    u_2(t) = \left(\frac{2\pi G \tilde{m}}{P}\right)^{1/3} \sin{i} \sin{\phi(t)}.
\end{equation}

\noindent where $P$ is the orbital period, $i$ the inclination, and $\phi(t) = 2\pi t/P + \phi_0$ where $t$ is the current time and $\phi_0$ is the orbital phase. For a position angle on sky PA measured E of N, the astrometric positions are:

\begin{align}
\label{eq:astro}
   \Vec{x}(t) &= \left(\frac{G \tilde{m} P^2}{4\pi^2}\right)^{1/3} 
   \begin{bmatrix}
   -\cos{\rm PA} \sin{\phi(t)} - \cos{i} \sin{\rm PA} \cos{\phi(t)}\\ 
   \sin{\rm PA} \sin{\phi(t)} - \cos{i} \cos{\rm PA} \cos{\phi(t)}
   \end{bmatrix}.
\end{align}

\noindent With only observations of the secondary's motion, the measurable combination of masses is $\tilde{m} = M/(1+q)^3$, resulting in a factor of $8$ range in allowed total mass $M$. Assuming the hot dust emission is centered on the binary center of mass, a single measurement of the offset ($x$, $y$) provides a lower limit to the semi-major axis $a$ on sky. The astrometric offset should be large when the radial velocity offset is near maximum, as selected by \citet{eracleous2012}. A proper motion measurement can be compared with the radial velocity offset, and a second derivative of either quantity measures the orbital period $P$. Combining positions and proper motions with radial velocity measurements provides enough information to constrain an orbit.

If the hot dust emission is stationary (e.g., uniform or asymmetric around the circumbinary disk), $\Delta x(t) = x(t) + x_0$ would be the measured quantity, with $x_0$ a potentially constant offset of the dust emission. If instead the hot dust emission follows the motion of the secondary as in the geometric model above, then $\Delta x(t) \simeq -(0.5-1.5) (1+q) x(t)$. An unknown pre-factor would produce additional scatter by a factor of $\simeq 10$ in the inferred value of $M$, but with weak dependence on $q$.

\begin{deluxetable*}{lcccccccc}
\tablewidth{0pt} 
\tablenum{1}
\tablecaption{Some candidate SMBHB GRAVITY+ targets \label{tab:gplustargets}}
\tablehead{
\colhead{SDSS ID} & \colhead{$z$} & \colhead{$K$} & \colhead{$V$} & \colhead{$\theta_{a, 1}$ ($\mu$as)} & \colhead{$\Delta \phi$ (deg)} & \colhead{$a_{\rm min}$ (pc)} & \colhead{$a_{\rm max}$ (pc)} & \colhead{Ref.}
}
\startdata
SDSS J001224.02-102226.2 & 0.2287 & 13.7 & 17.1 & 27.0 & 0.28 & 0.10 & 0.43 & 1\\
SDSS J015530.01-085704.0 & 0.1648 & 12.7 & 16.8 & 35.0 & 0.37 & 0.08 & 0.33 & 1\\
SDSS J091928.69+143202.6 & 0.2072 & 14.5 & 17.6 & 29.1 & 0.31 & 0.07 & 0.30 & 1\\
SDSS J093844.45+005715.7 & 0.1707 & 13.8 & 17.2 & 34.0 & 0.36 & 0.07 & 0.29 & 1\\
SDSS J111230.90+181311.4 & 0.1952 & 14.5 & 18.4 & 30.5 & 0.32 & 0.04 & 0.19 & 2\\
SDSS J115158.90+122128.9 & 0.1697 & 14.5 & 17.9 & 34.2 & 0.36 & 0.05 & 0.21 & 1\\
SDSS J125142.28+240435.3 & 0.1887 & 14.0 & 17.6 & 31.4 & 0.33 & 0.06 & 0.26 & 1\\
SDSS J140251.19+263117.5 & 0.1877 & 12.5 & 16.9 & 31.5 & 0.33 & 0.09 & 0.37 & 1\\
SDSS J153705.95+005522.8 & 0.1365 & 13.5 & 17.3 & 40.9 & 0.43 & 0.05 & 0.22 & 2\\
SDSS J155654.47+253233.5 & 0.1645 & 13.9 & 18.0 & 35.0 & 0.37 & 0.04 & 0.19 & 1\\
\enddata
\tablecomments{\label{tab:gplus_targets}Targets are selected as those with $K < 15$, Dec. $> 30^\circ$, and $0.09 < z < 0.25$ from the offset radial velocity SMBHB candidates identified by \citet{runnoe2017} (1) and \citet{guo2019} (2). The estimated astrometric size $\theta_{a,1}$ is scaled to a semi-major axis of $0.1$ pc using angular diameter distances from the target redshifts. The phase signal is calculated according to \autoref{eq:dphase} assuming a Pa $\alpha$ line strength of $f_{\rm line} = 0.1$ and $q = 0.1$. The allowed range of semi-major axis for astrometric measurements is inferred from the optical luminosity as described in the text.}
\end{deluxetable*}

\section{Astrometric measurements with GRAVITY+}
\label{sec:gplus}

Currently known candidate SMBHBs with single, offset, moving broad emission lines are generally found at $z \simeq 0.2$, with apparent magnitudes of $V \lesssim 18$ and $K \lesssim 15$ \citep{runnoe2017,guo2019}. For a semi-major axis of $a \simeq 0.1$ pc, the size on sky $\theta_a \simeq 30 \, \mu$as, while the BLR size is a factor of several smaller. These properties are well matched to the expected sensitivity of the planned upgrade to the GRAVITY instrument, GRAVITY+. Through a combination of ongoing and near future upgrades including new grisms, improved VLTI vibration control, new AO systems, and laser guide stars the goal is to reach limiting magnitudes $K \lesssim 14-15$ with comparable astrometric accuracy as is currently possible for $K \lesssim 10-11$.\footnote{See \url{https://www.mpe.mpg.de/ir/gravityplus} for more details.}

\subsection{Differential phase astrometry}

The astrometric offset of an emission line of strength $1+f$ relative to the normalized hot dust continuum is measured by the differential phase $\Delta \phi = \phi(\lambda) - \phi_c$,

\begin{align}
\label{eq:dphase}
    \Delta \phi &= -2\pi \frac{f}{1+f} (u \Delta x + v \Delta y) \\
    |\Delta \phi| & \simeq 0.3^\circ \left(\frac{f_{\rm line}}{0.1}\right) \left(\frac{a}{0.1 \, \rm pc}\right) \left(\frac{D_A(z)}{500 \, \rm Mpc}\right)^{-1} \left(1+q\right)^{-1},
\end{align}

\noindent with $\Delta x(t)$ and $\Delta y(t)$ the astrometric offsets discussed above. The line strength $f$ is normalized to the continuum flux, $f_{\rm line} = f/(1+f)$, and $D_A(z)$ is the angular diameter distance. The differential phase signal of a wavelength-independent $(x,y)$ offset has the shape of the emission line itself, with an amplitude depending on the ($u$,$v$) coordinates of each baseline.

\begin{figure*}
\begin{tabular}{lll}
\includegraphics[width=0.48\textwidth]{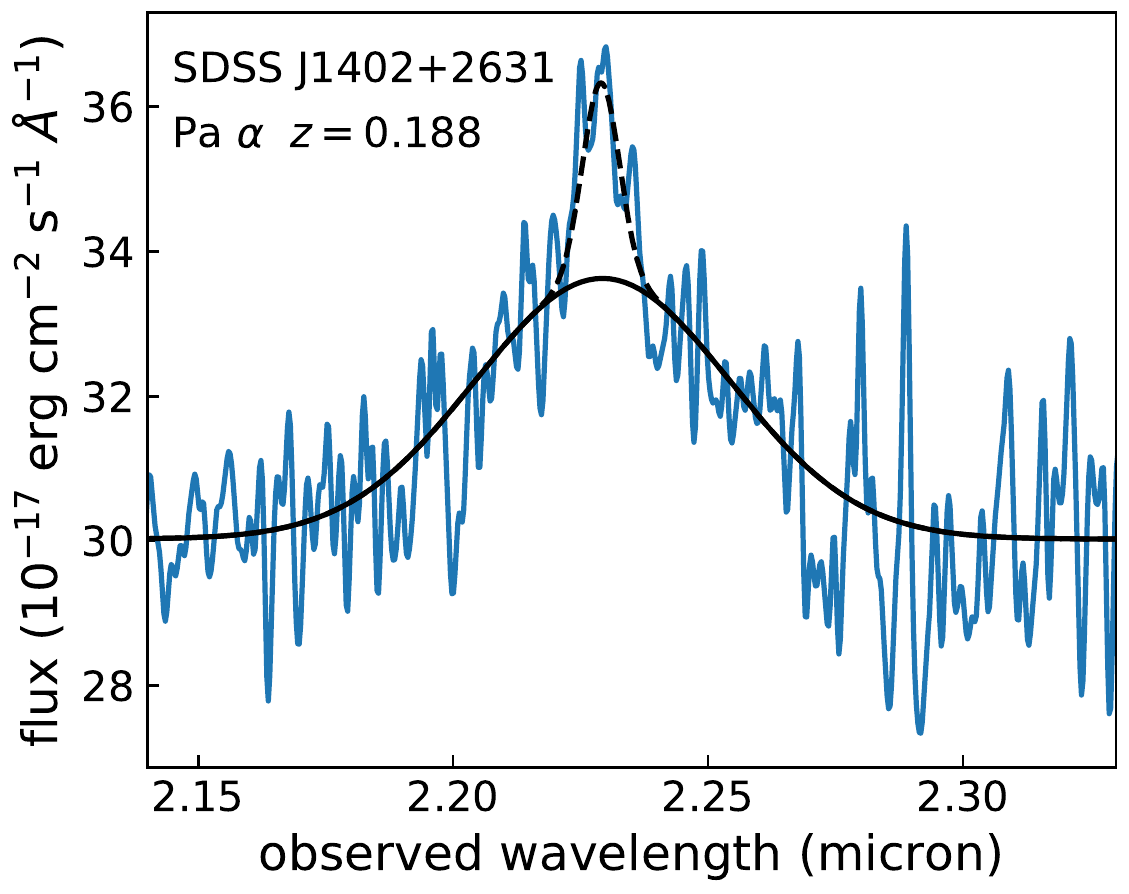} &
\includegraphics[width=0.52\textwidth]{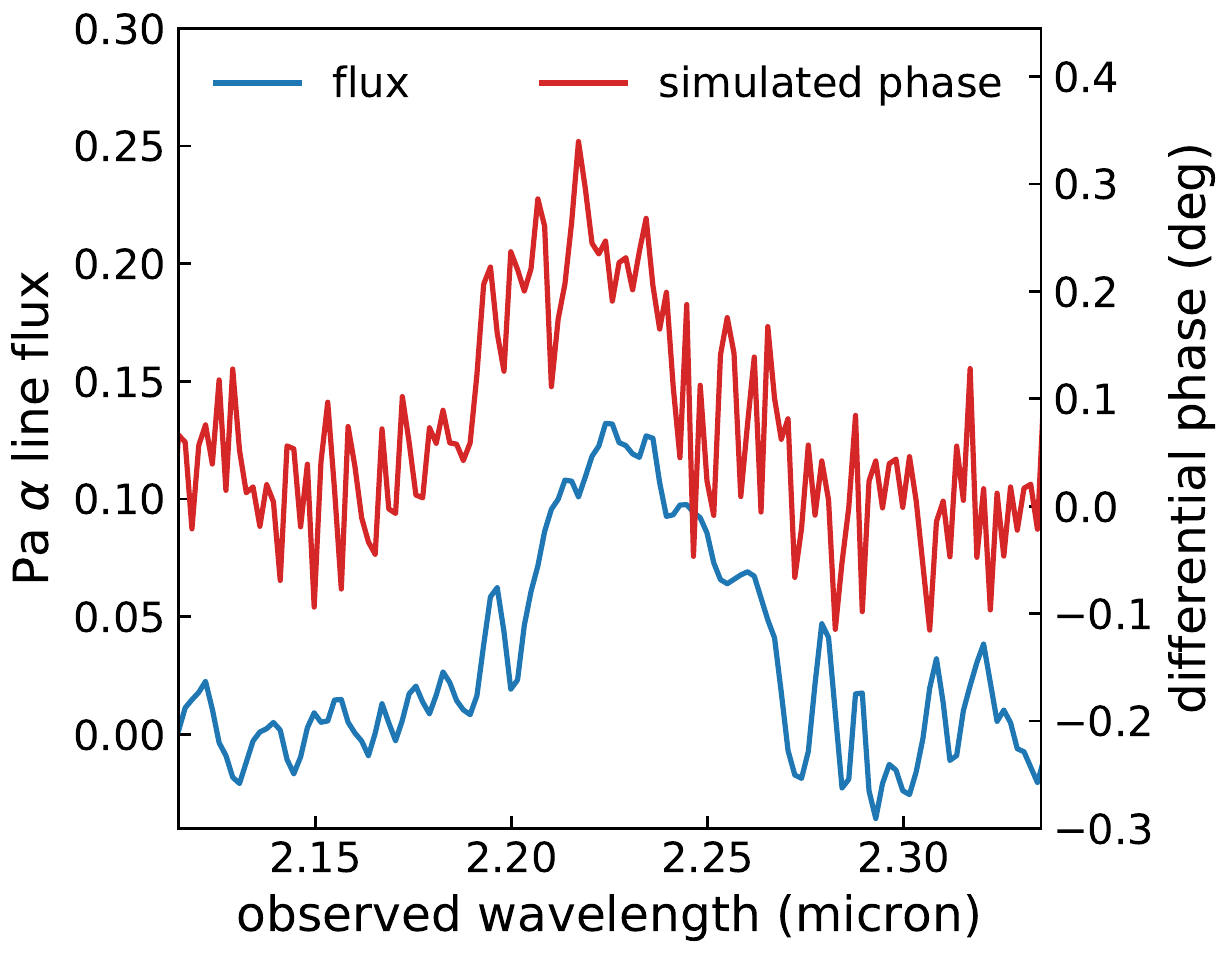}\\
\includegraphics[width=0.48\textwidth]{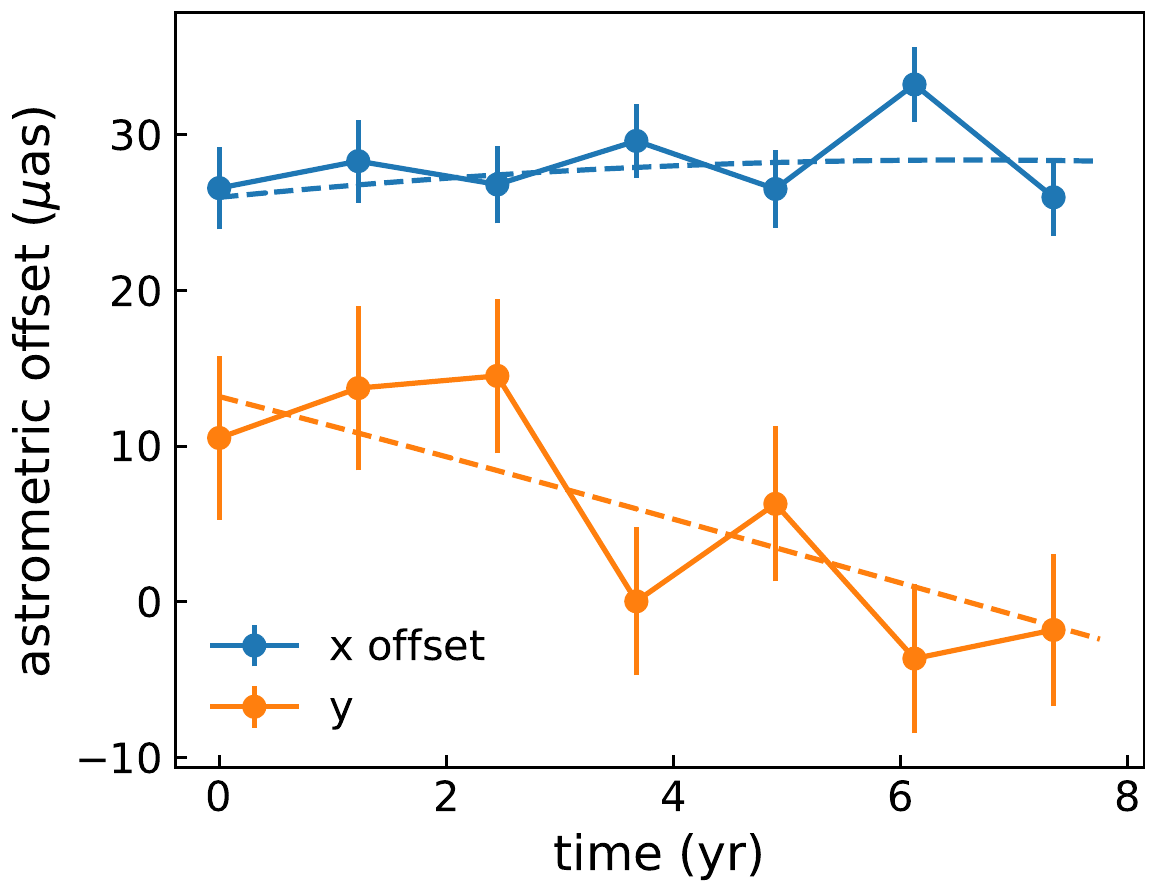}
\end{tabular}
\caption{Measured Pa $\alpha$ line profile of SDSS J140251.19+263117.5 (left), showing narrow (dashed) and broad (solid) components. The broad component velocity width is $\sigma \simeq 3300 \, \rm km \, \rm s^{-1}$. We used the broad line profile component model to simulate differential phase data (right) corresponding to our fiducial orbital parameters of $\tilde{m} = 10^9 M_\sun$, $P = 100$ yr, and $i = 25^\circ$ and assuming the continuum photocenter is stationary at the center of mass. The assumed phase error is $0.1^\circ$ per VLTI baseline, resulting in astrometric errors of $\simeq 2$ and $4 \, \mu$as in RA and Dec (bottom).}
\label{fig:gplus_phase}
\end{figure*}

\begin{figure*}[t]
\begin{tabular}{cc}
\includegraphics[width=0.45\textwidth]{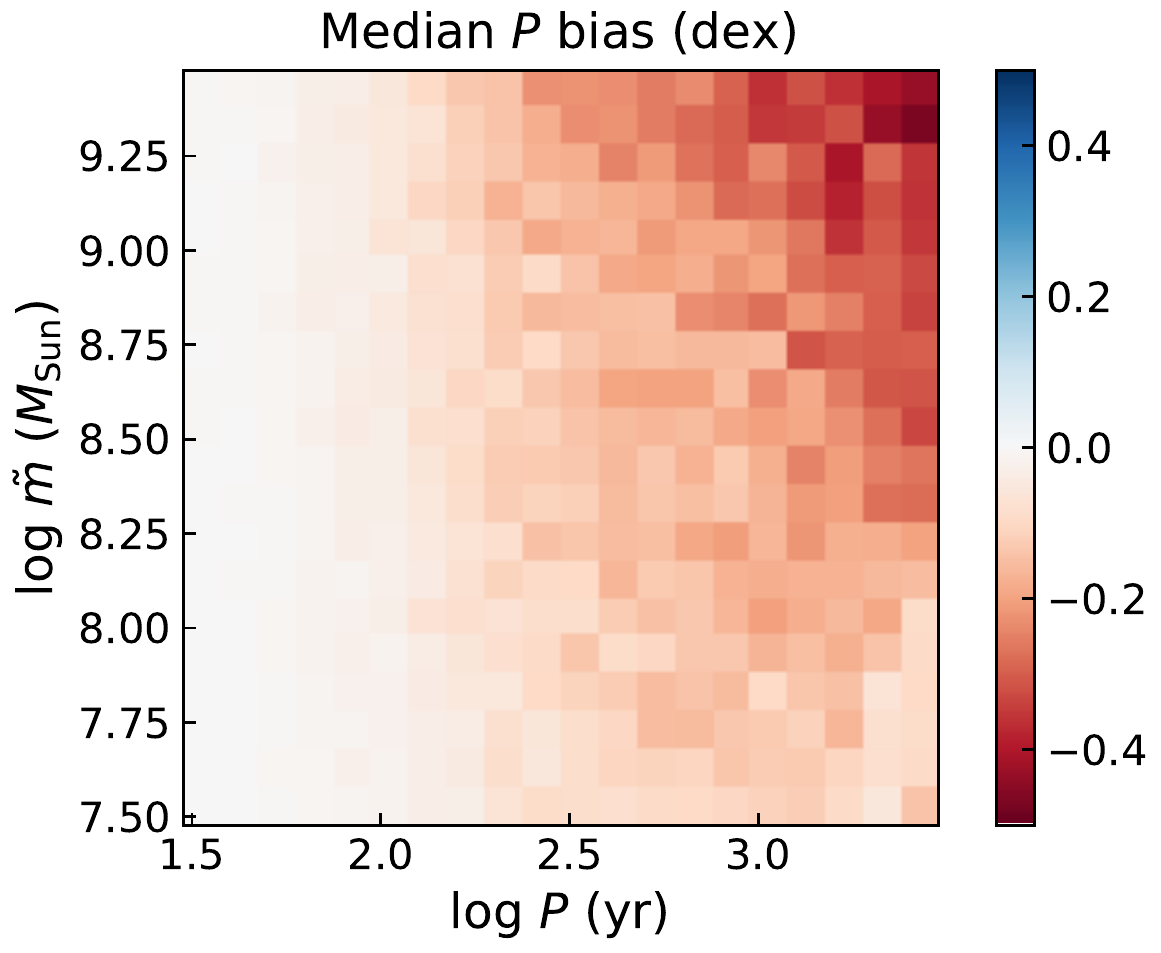} &
\includegraphics[width=0.45\textwidth]{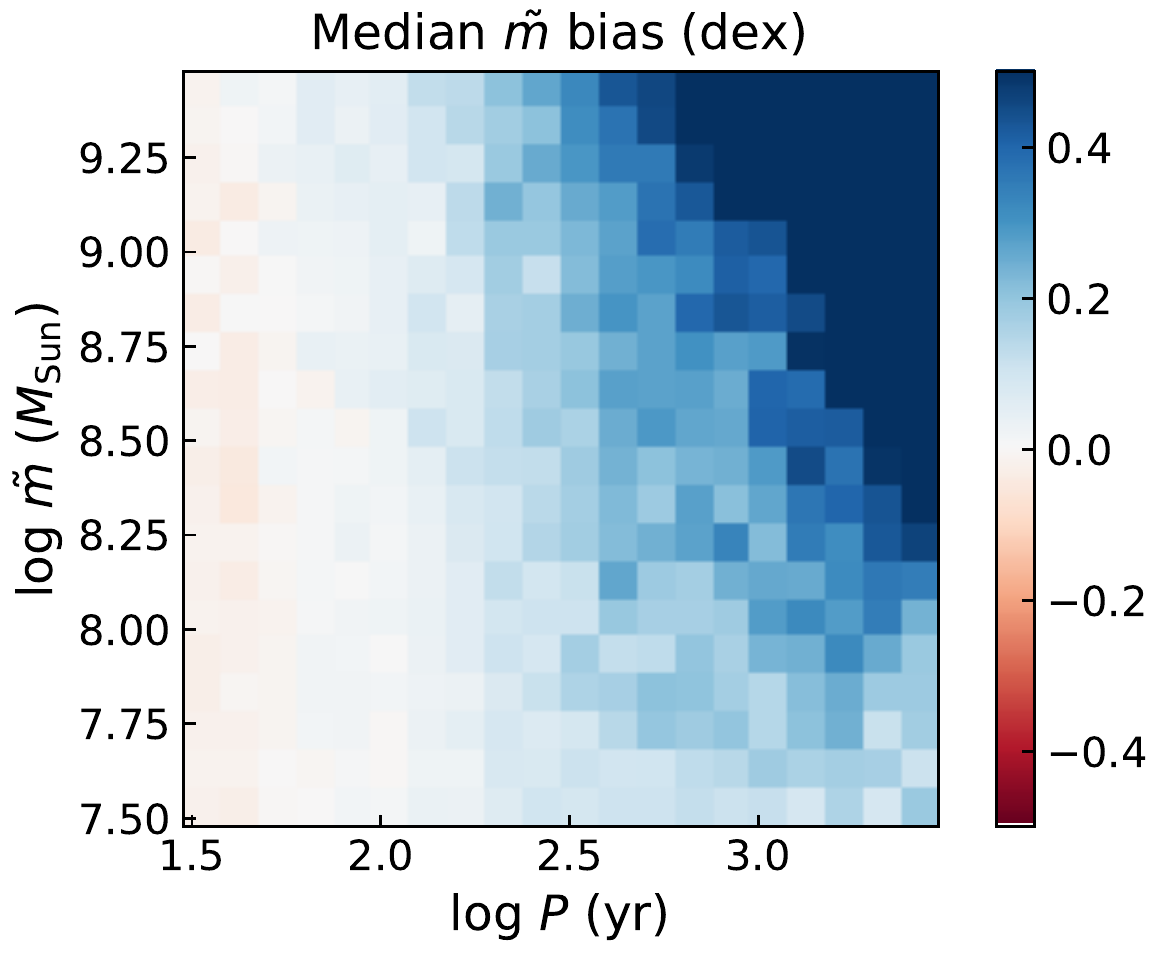}\\
\includegraphics[width=0.45\textwidth]{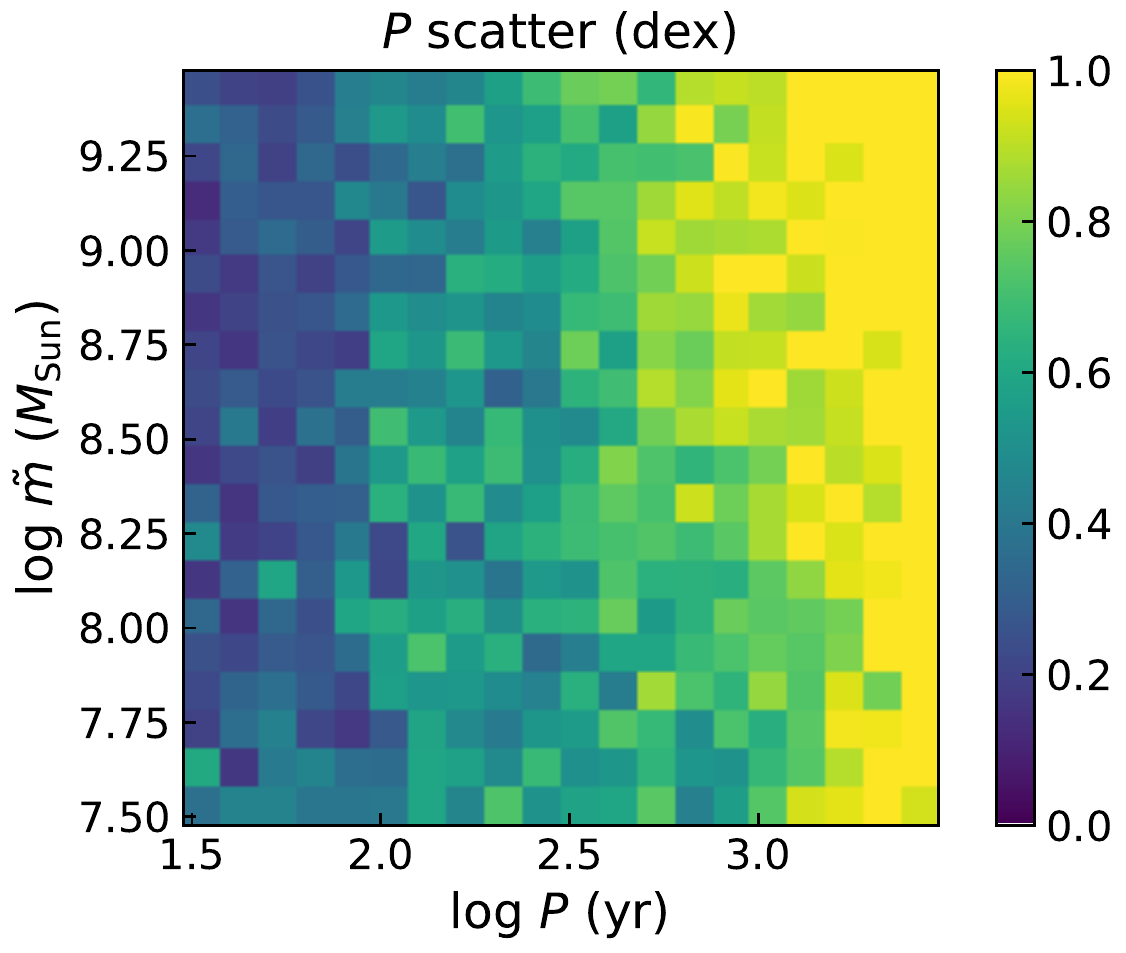} &
\includegraphics[width=0.45\textwidth]{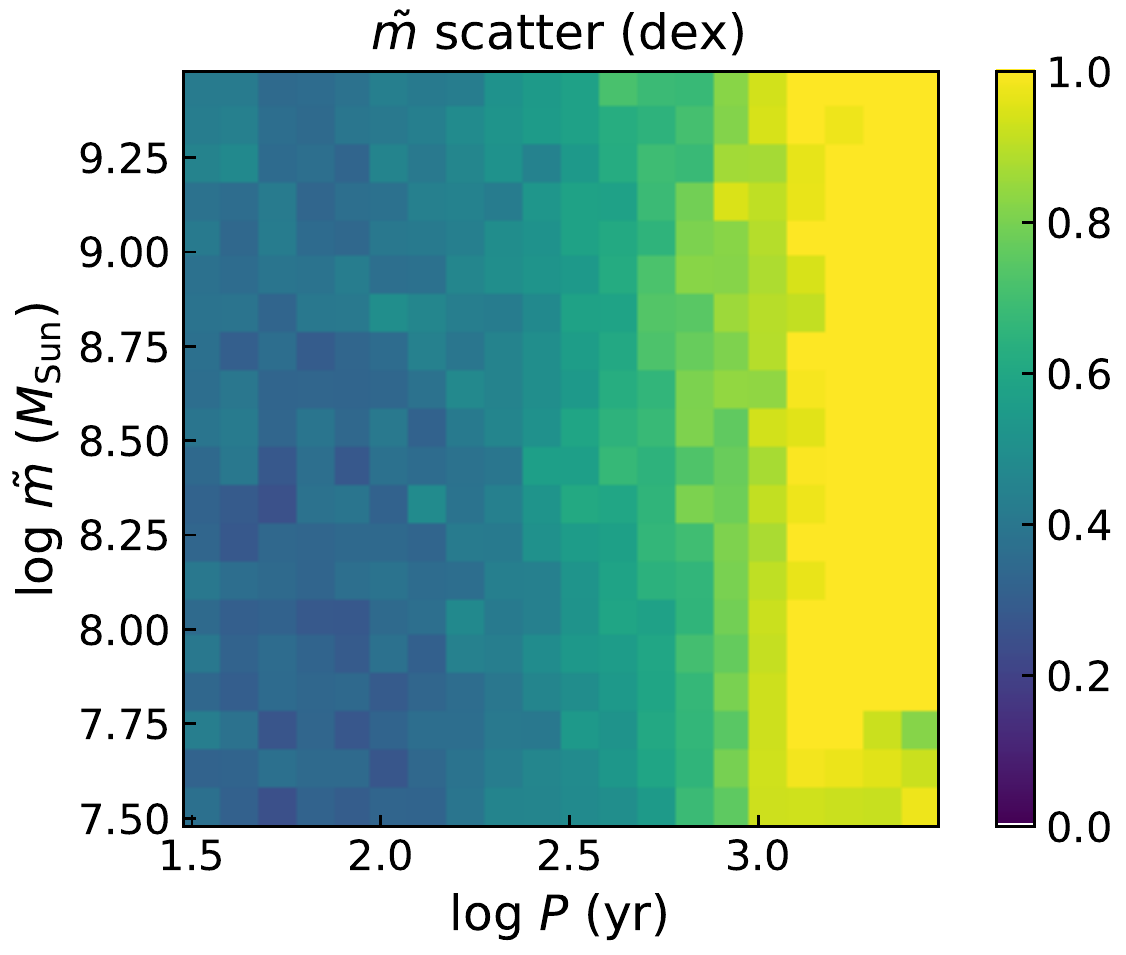}
\end{tabular}
\caption{Bias (top) and scatter (bottom) in median inferred values of $P$ (left) and $\tilde{m}$ (right) from simulated astrometric and radial velocity measurements as a function of those parameters. The inferred values of $P$ ($\tilde{m}$) are underestimates (overestimates) when both the period and mass are large. For periods $\lesssim 10^3$ yr, we reliably recover both parameters with scatter $\lesssim 0.6$ dex.}
\label{fig:fit_grid}
\end{figure*}

\subsection{A case study with SDSS J1402+2631}

From the parent radial velocity samples of \citet{runnoe2017} and \citet{guo2019}, we have listed properties of some SMBHB candidate targets visible from the VLTI (Dec $< 30^\circ$) with $K < 15$ and $0.09 < z < 0.25$ in \autoref{tab:gplustargets}. For those redshifts, the Pa $\alpha$ line is redshifted into the GRAVITY K band. All $10$ targets have predicted phase signatures of $\gtrsim 0.3^\circ$ for a $0.1$ pc binary orbit. As such they form a promising first set of candidates for GRAVITY+ astrometry.

We have further used the observed optical luminosity, radial velocity offset, and minimum periods for the sample to constrain the parameter space where astrometric monitoring might be feasible. Following \autoref{sec:param_space}, we calculate the allowed range of semi-major axis from $R_{\rm BLR} = R_L(q,a_{\rm min})$ and $a_{\rm max} = 2 R_{\rm dust}$. As shown in \autoref{tab:gplus_targets}, we are sensitive to binary semi-major axes of $\simeq 0.05-0.4$ pc. This range depends on the mass ratio $q$, in the sense that $a_{\rm min}$ increases with decreasing $q$. The full range is feasible for nearly equal mass binaries with $q \lesssim 1$. We can impose further constraints to estimate allowed total binary mass ranges. We require a total binary mass that 1) results in $P_{\rm min} < P < 10^3$ yr, where $P_{\rm min}$ is the minimum period obtained from fitting the measured radial velocity curves \citep{runnoe2017}, 2) can match the observed radial velocity offset $u_2$ (\autoref{eq:rv}), and 3) results in an Eddington ratio of $10^{-3} < L_2 / L_{\rm Edd} < 3$ for the secondary. All of those constraints are satisfied for total masses of $M \sim 10^{7-10} M_\sun$.

As one example, we consider the object SDSS J1402+2631. We have measured the Pa $\alpha$ emission line profile of this quasar (\autoref{fig:gplus_phase}) using the TripleSpec instrument at the Apache Point Observatory 3.5m telescope. Observations were taken in June 2020 with the $1.1\arcsec$ slit in a standard nodding ABBA sequence of $8\times120$s exposures. The seeing was $1\arcsec$. The data were reduced using a modified version of the \texttt{Spextool} package \citep{cushing2004}, and an A0V star was used for telluric correction \citep{vacca2003}. We detect broad emission lines of Pa $\alpha$, $\beta$, $\gamma$, $\delta$, $\epsilon$ at a redshift of $z \simeq 0.188$ in the JHK band spectra. The continuum flux corresponds to $K = 12.8$, similar to the $K = 12.5$ measured by 2MASS. \autoref{fig:gplus_phase} shows a decomposition of the Pa $\alpha$ emission line into Gaussian broad and narrow components, where the broad line component has a velocity width $\sigma \simeq 3300 \, \rm km \, \rm s^{-1}$ and peak relative line strength of $f \simeq 0.12$. The line width is consistent with the reported range of H$\beta$ FWHM \citep{runnoe2015}.

We use the broad line component model to simulate GRAVITY+ data, adopting a phase error of $0.1^\circ$ per baseline as achieved in observations of bright ($K \sim 10-11$) AGN to date with GRAVITY \citep[][]{gravity2018,gravity2020iras}. We take VLTI ($u$,$v$) coordinates of this Northern target from \texttt{Aspro} \citep{aspro}. The top right panel of \autoref{fig:gplus_phase} compares the measured line profile and simulated differential phase signals for fiducial parameters of $\tilde{m} = 10^9 \, M_{\rm sun}$, $a = 0.1$ pc, $P = 100$ yr, using the model described in \autoref{eq:rv} and \autoref{eq:astro} and assuming a stationary continuum photocenter. The differential phase is averaged over the 3 longest baselines. Fitting \autoref{eq:dphase} for the offset ($x$,$y$) results in errors of $\simeq 2\times4 \, \mu$as. The measured offsets and errors are shown compared to the underlying model in the bottom panel of \autoref{fig:gplus_phase}. Both proper motion and acceleration would be detected from astrometric monitoring, resulting in confirmation of the target as an SMBHB and allowing estimates of $\tilde{m}$ and $P$, in combination with radial velocity measurements. The parameters $\phi_0$ and PA are  difficult to constrain, likely due to the low inclination angle.

\subsection{Mass and period estimates from a parameter survey}
\label{sec:fitting}

We next perform a mock parameter survey to see how well binary mass and period information might be recovered. We consider periods of $3 \times 10^{2-4}$ yr and $\tilde{m} = 3 \times 10^{7-9} M_\sun$. We generate 10 epochs of simulated radial velocity data taken over a $25$ year time baseline (since current candidates have $\lesssim 15$ year time baselines) with errors of $100 \, \rm km \, \rm s^{-1}$ intended to mimic the ``jitter" noise which dominates the error budget in many current candidates \citep{runnoe2017}. We generate $10$ epochs of astrometric data over $8$ years, adopting errors of $4 \, \mu$as in both the $x$ and $y$ (RA and Dec) coordinates.

For each combination of $P$ and $\tilde{m}$, we generate $N = 300$ realizations of mock data, varying the random error realization as well as the parameters of $i$, $\phi_0$, and PA. The inclination is constrained to be $i < 75\deg$, while $\phi_0$ and PA are varied over their full ranges. We use a least squares method to identify the best fitting parameters in each case. The initial guess for least squares is fixed to fiducial values of $\tilde{m} = 10^{8.5} \, M_\sun$ and $P = 300$ yr. The median parameter bias and scatter over the $N = 300$ simulations for each parameter combination are shown in \autoref{fig:fit_grid}, excluding the $\simeq 2\%$ of simulations where the minimization method fails. We recover the input parameters with errors of $\lesssim 0.6$ dex for periods of $P \lesssim 10^3$ yr. For longer periods, second derivatives are usually not detected in radial velocity or astrometry. For $P \gtrsim 300$ yr and $\tilde{m} \gtrsim 10^8 M_\sun$, the recovered parameters show bias, in that they systematically find shorter periods and smaller $\tilde{m}$ than the input values. 

\begin{figure*}[t]
\begin{tabular}{cc}
\includegraphics[width=0.47\textwidth]{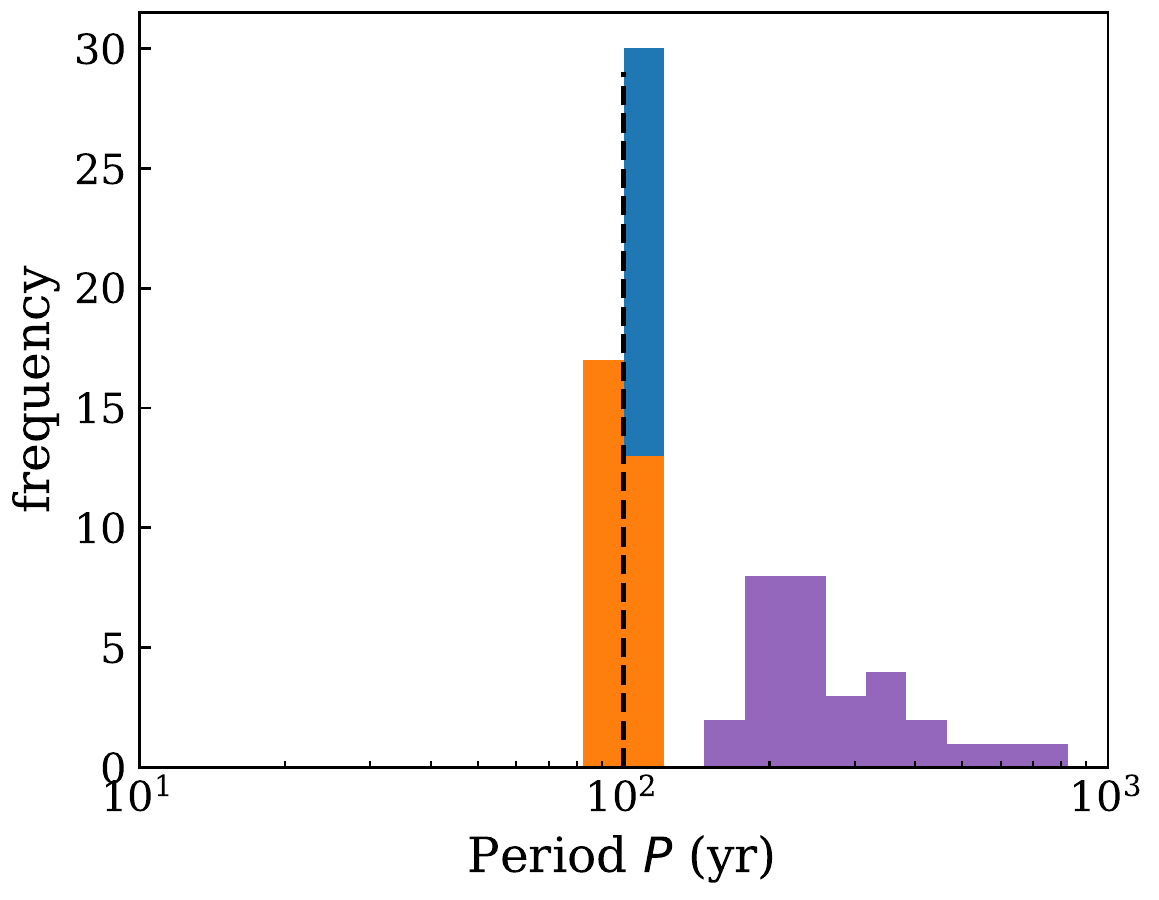} &
\includegraphics[width=0.47\textwidth]{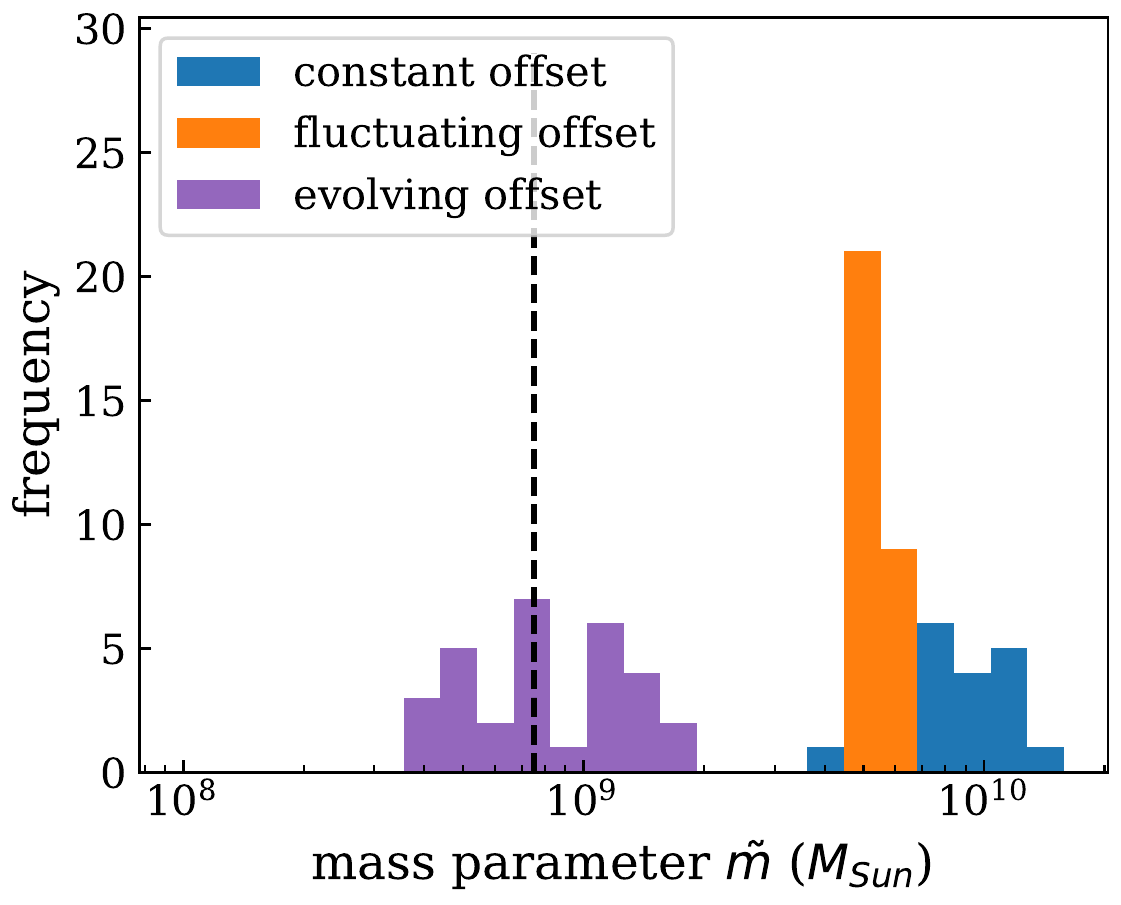}
\end{tabular}
\caption{Distributions of inferred values of $P$ (left) and $\tilde{m}$ when fitting the pure orbital model (\autoref{eq:astro}) to simulated data using i) a time-independent, offset hot dust continuum, ii) a fluctuating hot dust continuum offset, and iii) the evolving continuum model where the offset tracks the secondary's orbital motion. We recover proper motions and accelerations in all cases. Biases in $\tilde{m}$ or $P$ are generally introduced in the fluctuating and evolving continuum cases, where the inferred values are incorrect due to the use of the wrong hot dust emission model.}
\label{fig:errors}
\end{figure*}

\section{Discussion}
\label{sec:discussion}

Current sub-pc SMBHB candidates with single, moving, broad emission lines have $K \sim 12-15$ and sizes on sky of $\theta_a \simeq 30 \, \mu$as ($a / 0.1$ pc). It may be possible to trace the binary orbit in these systems with the upgraded near-infrared interferometry instrument GRAVITY+ at the VLTI. A monitoring campaign over $5-10$ years could reveal proper motions and accelerations, resulting in robust detections of the progenitors of merging supermassive black holes and constraining their system parameters.

As an example, using the Pa $\alpha$ profile of one current candidate and current GRAVITY phase noise we find astrometric errors of $\lesssim 4 \, \mu$as. We simulate a combined radial velocity and astrometric campaign, resulting in robust detections of binaries with $\lesssim 0.5$ dex measurements of $\tilde{m}$ and $P$ for systems with $P \lesssim 10^3$ yr where accelerations can be measured. 

With radial velocity data alone, generally $P$ can still be well constrained, since radial velocity changes (accelerations) can usually be measured over our assumed $25$ years of monitoring. Constraining $\tilde{m}$ requires astrometry. We also note that the complicating issues of line profile changes and jitter noise would not impact the astrometric offset measurement. The differential phase signal is proportional to the ratio of line to total flux, even for a variable line profile.

In principle, combining astrometric and radial velocity data we can fit for the angular diameter distance $D_A$ without using the redshift. The result would then provide a cosmological constraint. As expected, fitting directly for the distance results in a strong correlation between $\tilde{m}$ and $D_A$, while $P$ remains well measured. In our tests, precise measurements of both $\tilde{m}$ and $D_A$ require astrometric errors of $\lesssim 1 \, \mu$as and/or astrometric campaigns of $\gtrsim 25$ yr. This may be feasible for short period systems, and/or if even higher astrometric precision becomes possible.

The differential phase measurement is referenced to the continuum photocenter position. The continuum near-infrared emission is due to hot dust, whose origin and time evolution in the SMBHB scenario is unclear. We have considered two extreme cases. In one case, the continuum is stationary with a photocenter at the center of mass of the binary. In this case, relative astrometry directly measures the orbital position of the secondary black hole. We have used this model to generate synthetic data above. 

We have considered a simple geometric model of the second case, assuming that the hot dust emission originates in the circumbinary disk at the distance of the sublimation radius away from the secondary. In this case, the hot dust photocenter tracks the orbital motion of the binary. Remarkably, over a large portion of the relevant parameter space (\autoref{fig:paramspace}) the relative offset in this model is opposite in sign and comparable in amplitude to the orbital motion of the secondary (\autoref{fig:offset_circle}). In the evolving continuum model, it is possible in principle to measure both $\tilde{m}$ and $q$, e.g. the two black hole masses $M_1$ and $M_2$. This seems to require lower measurement errors and longer campaigns than we have assumed.

A time variable central luminosity will produce fluctuations of the hot dust photocenter due to differential light travel time delays \citep[reverberation,][]{shen2012,dorazio2017}. For relatively small fluctuations, the maximum amplitude of this effect has comparable contributions from changes in the hot dust emission radius and intensity ($\Delta x / a \lesssim 10\%$ each for $\Delta L / L \simeq 20\%$ at $i = 30^\circ$). We evaluate the possible impact of uncertainties in the hot dust structure and its time variability using experiments with fake data. We consider models with i) a constant hot dust offset (e.g. due to asymmetry), ii) a fluctuating hot dust offset due to luminosity variations of $\Delta L / L \simeq 20\%$ using a measured $R$ band light curve of 3C 273 \citep{fan2014}, and iii) an evolving offset tracking the orbit according to the geometric model described above. In each case, we run $30$ trials of fitting the static dust orbital model (with no continuum photocenter offset, \autoref{eq:astro}) to the generated data and errors. Data are generated with $\tilde{m} = 10^{8.5} M_\sun$ and $P = 100$ yr, and errors of $100 \, \rm km \, \rm s^{-1}$ in radial velocity and $4 \mu$as in astrometry. As in \autoref{sec:fitting}, we identify the best fitting parameters using a least squares method. Distributions of the identified best-fitting $P$ and $\tilde{m}$ are shown in \autoref{fig:errors}. For the constant and fluctuating offset cases, the mass parameter is overestimated. Depending on the choice of parameters, we have also found underestimates. For the evolving offset case, the mass parameter is well recovered while the orbital period is overestimated. These biases are introduced by the use of an incorrect hot dust emission model. In all cases, proper motions and accelerations can still be detected. 

In the evolving dust scenario, the hot dust emission region size is smaller and concentrated on one side of the circumbinary disk. The amplitude of the reverberation offset will be smaller as a result. However, the light travel time delay will cause the offset vector $\Delta \vec{x}$ between the BLR and hot dust to point slightly away from the center of mass. In principle, current GRAVITY observations could detect both the fluctuating sublimation radius size and reverberation effect using differential amplitude and phase data \citep[e.g.,][]{gravity2020dust} from different epochs where the continuum luminosity varies.

The same interferometry measurements proposed here could help distinguish scenarios for the hot dust continuum emission and its time variability in SMBHB candidates. The evolving continuum model would generically predict a smaller size than the stationary dust model for $R_{\rm sub} \lesssim 2a$. Candidates in that regime should show larger (smaller) dust sizes than expected from the radius-luminosity relation according to the stationary (evolving) dust emission models. The evolving continuum model might also show time-variable, asymmetric structure. Further constraints on both hot dust and BLR evolution would be possible if more distant, narrow emission line components of Pa $\alpha$ or Si $[$VI$]$ are present, since they could be used as independent, static phase references. 

We have focused on targets with offset, moving broad emission lines and assumed that the broad emission line originates from atomic gas centered on the secondary black hole. In the model of \citet{nguyen2020}, the larger BLR size around the primary could result in substantial contributions from its own line flux. If most of the atomic line emission is from around the primary, the astrometric signals considered here will be suppressed by a factor of $q/(1+q)$, and interferometry measurements would be most sensitive to large mass ratios of $q \gtrsim 1/3$. Our simulations have also used circular binary orbits. The same measurements are in principle possible if the binary is driven to high eccentricity. Additional time variability of the accretion luminosity and circumbinary disk size and shape could result in larger fluctuations of the hot dust photocenter location in this case.

Photometric candidates showing sinusoidal optical variations \citep[e.g., PG 1302$-$102,][]{graham2015} should also be sufficiently bright to detect with GRAVITY+. For the very short periods $\lesssim 30$ yr accessible with photometric data to date, a single complex BLR structure might surround both black holes \citep[e.g.,][]{shen2010}. The astrometric signature in that case is unclear. \citet{songsheng2019} calculated velocity-dependent photocenter signatures of a binary system with two active black holes, each with its own BLR. They further assumed a continuum photocenter at the center of mass, and identical Eddington ratios for both black holes. Relaxing either of those assumptions \citep[e.g.,][]{dorazio2019} would result in an additional velocity-independent astrometric offset like that discussed here. \citet{kovacevic2020} presented a first exploration of the combined effects for a somewhat different parameter regime than explored here. Both an overall offset of the hot dust and BLR photocenters, and velocity-resolved kinematics of the BLR have been detected recently in IRAS 09149$-$6206 \citep{gravity2020iras}, providing independent measurements of the photocenter offset and BLR size.

We identified $10$ possible candidates, which show evolution in radial velocity consistent with binary motion in $\ge 3-5$ epochs over $5-15$ yr \citep{runnoe2017,guo2019}. Large spectroscopic surveys will likely add additional candidates in the next several years. For example, the SDSS-V Black Hole Mapper program plans to take between $3-13$ spectra of each of $25000$ quasars \citep{sdssv}. Additional candidates in the southern sky would be particularly promising for GRAVITY+ observations, since deep integrations of $\simeq 4-8$h may be required to achieve the astrometric accuracy needed to confirm candidate systems as SMBHBs and map out their orbits.

\acknowledgments
JD thanks C. Gray, A. Kowalski, K. Davis, D. Dewitt, M. Jacobs, and J. Crowley for their help in obtaining the APO data used here. We thank T. Bogdanovic, D. D'Orazio, S. H\"{o}nig, Y. Shen, J. Runnoe, and the anonymous referee for helpful comments which improved this manuscript. JD was supported in part by NSF grant AST-1909711 and an Alfred P. Sloan Research Fellowship.

\appendix

\section{Astrometric offset for the geometric evolving continuum model}
\label{app:circles}

Consider two circles, one describing the inner edge of the circumbinary disk of radius $2a$ centered on the center of mass, and one with radius $R_{\rm sub}$ centered on the secondary black hole (the ``sublimation ring"), offset (without loss of generality) in the $-x$ direction by a distance $a/(1+q)$ (see right panel of \autoref{fig:circle_diagram}). When the two circles intersect, we calculate the offset between the line and continuum emission as the centroid of the arc of the sublimation ring which intersects the circumbinary disk. The centroid of the arc is,

\begin{equation}
    \Delta x = -R_{\rm sub} \frac{\sin{\alpha}}{\alpha},
\end{equation}

\noindent where the offset $\Delta \vec{x} = \vec{x}_{\rm BLR} - \vec{x}_{\rm dust}$ is negative, and $\alpha$ is the half-angle of the arc,

\begin{align}
    \sin{\alpha} &= \frac{a(1+q)}{2 R_{\rm sub}} \xi(q,R_{\rm sub}/a),\\
    \xi(q,r) &= \sqrt{(q_o^2-r^2)(r^2-q_i^2)},
\end{align}

\noindent where $q_i = (1+2q)/(1+q)$ and $q_o = (3+2q)/(1+q)$ bound the range of solutions where the two circles intersect. To calculate the correct half-angle, we need to switch solutions at a transition point $r_t = \sqrt{q_i q_o}$ given by $\partial[\xi(q,r)/2r]/\partial r = 0$, where $r = R_{\rm sub}/a$. A piecewise expression for the offset is then,

\begin{equation}
    \Delta x = \begin{cases}
    -a r & r < q_i\\
            \displaystyle
    -a(1+q) \frac{\xi(q,r)/2}{\arcsin{\left[(1+q)\xi(q,r)/2r\right]}} & q_i \le r \le \sqrt{q_i q_o}\\
            \displaystyle
        -a(1+q) \frac{\xi(q,r)/2}{\pi-\arcsin{\left[(1+q)\xi(q,r)/2r\right]}} & \sqrt{q_i q_o} < r \le q_o\\
        0 & r > q_o.
        \end{cases}
\end{equation}

\noindent The expression can be written more compactly using $\arcsin z = z R_C(1-z^2,1)$ for $-1 \le z \le 1$, where $R_C(x,y)$ is the \citet{carlson1979} circular function:

\begin{equation}
    \Delta x = \begin{cases}
    -a r & r < q_i\\
        \displaystyle
    -\frac{a r}{R_C (1-z^2,1)} & q_i \le r \le \sqrt{q_i q_o}\\
            \displaystyle
        -\frac{a r z}{\pi - z R_C(1-z^2,1)} & \sqrt{q_i q_o} < r \le q_o\\
        0 & r > q_o,
        \end{cases}
\end{equation}

\noindent and $z = (1+q) \, \xi(q,r) / 2r$. 

When $r = q_i$, $z = 0$ and $R_C(1,1)=1$ so that $\Delta x = -a r$. When $r^2 = q_i q_0$, $z = 1$ and $R_C(0,1) = \pi/2$ and the solutions again match on smoothly. When $r = q_0$, $z = 0$ and $\Delta x = 0$. We have verified the expressions for the arc centroid through comparison with a direct numerical calculation using discretized circles.

\bibliographystyle{aasjournal}

\end{document}